\documentclass[preprint,showpacs,preprintnumbers,amsmath,amssymb, superscriptaddress]{revtex4}

\usepackage{graphicx}
\usepackage{dcolumn}
\usepackage{bm}
\usepackage{color}

\begin{document}

\preprint{APS/123-QED}

\title{
Reproduction of a Protocell by Replication of Minority Molecule in Catalytic Reaction Network
}

\author{Atsushi Kamimura}
 \affiliation{Department of Applied Physics, Graduate School of Engineering,
 The University of Tokyo, 7-3-1, Hongo, Bunkyo-ku, Tokyo 113-8656, Japan.}
\author{Kunihiko Kaneko}
\affiliation{Department of Basic Science, Graduate School of Arts and Sciences,
The University of Tokyo,
3-8-1, Komaba, Meguro-ku, Tokyo 153-8902, Japan}
\affiliation{Complex Systems Biology Project, ERATO, JST, Tokyo, Japan.}
\date{\today}

\begin{abstract}
For understanding the origin of life, it is essential to explain the development of a
compartmentalized structure, which undergoes growth and division, from a set of chemical reactions.
In this study, a hypercycle with two chemicals that mutually catalyze each other is considered
in order to show that the reproduction of a protocell with a growth-division process
naturally occurs when the replication speed of one chemical is
considerably slower than that of the other chemical.
It is observed that the protocell divides after a minority molecule is replicated at a slow synthesis rate, and thus, a synchrony between the reproduction
of a cell and molecule replication is achieved.
The robustness of such protocells against the invasion of parasitic molecules is also demonstrated.
\end{abstract}

\maketitle

Filling a gap between just a set of catalytic reactions and
a reproducing cell is a difficult problem;
however, it is essential to fully understand this relationship to unveil the origin of life,
to establish a theory for a living state,
and to experimentally synthesize protocells\cite{protocell}.
Several theoretical attempts to understand the relationship have been made by using catalytic networks\cite{Dyson,Hypercycle,Eigen2,models}.
For reproduction of cells, it is necessary to replicate functional polymers.
At a primitive stage of life, however, errors in the sequence of
polymers are inevitable during their replications, which prevent
the transfer of information through
replications. This was pointed out by Eigen and Schuster
as an error catastrophe\cite{Hypercycle}.
To resolve this error-catastrophe problem,
they\cite{Hypercycle} proposed a hypercycle in which different
molecular species mutually catalyzed the replication of each other.
This hypercycle, however, introduced another problem, i.e.,
it led to the emergence of parasitic molecules that are catalyzed by other molecules
but do not in turn catalyze other molecules. Because such parasitic molecules
are in a majority, the reproduction of mutually catalytic molecules cannot be
sustained.

One possible way to resolve the problem of parasitism is to use
compartmentalization\cite{Eigen2, MaynardSmith, Szathmary}.
When molecules are compartmentalized in a protocell, the unit of selection is shifted from a molecule to a protocell. Then,
the protocells dominated by parasites are selected out, leaving behind protocells
without parasites.
However, such a compartmentalization introduces further complexity
to the primitive form of life.
It might be beneficial to use other molecules that act as a membrane; 
however, these molecules also need to be replicated.
A specific reaction-diffusion mechanism to create compartments is 
also proposed\cite{McCaskill, Hogeweg}.
To avoid the further introduction of complexity,
it is important to develop a simple mechanism for compartmentalization of a protocell. 

Another important issue associated with protocells 
 is the origin of heredity, i.e.,
the existence of specific molecules (such as DNA in the present cell) that are well preserved in the
offspring cells and also control the behavior of the cells.
By considering a simple hypercycle with only two chemicals, we have previously shown that
the chemical with a slower synthesis rate and (as a result) that in minority
plays the role of a heredity-carrier molecule. In fact,
the relevance of such minority molecules in the hypercycle for evolvability and for the elimination of
parasites is demonstrated\cite{KanekoYomo,Kaneko}.
Here again, reaction in a compartment and the division of such a protocell was assumed.

Here, we address the following questions. Can we explain
the origin of a protocell with compartmentalization and
its growth-division cycle, in terms of a reaction-diffusion process with
minority molecules? How does the division process of protocells maintain
synchrony with the replication of such minority molecules (i.e., what is the relationship
between cell division and replication of DNA?)?
We address these questions by simply introducing spatial diffusion of molecules and catalytic reactions.
We use the physicists' golden rule to study one the most simplest systems, i.e.,
a system consisting of only two chemical species that mutually catalyze the replication of each other,
with the Brownian motion of molecules. From the results of a stochastic simulation, we show that the
difference between the synthesis speed of the two molecular species leads not only to the formation of
compartments as a cluster of such molecules
but also to their recursive division with a certain cell size, even though no complicated mechanisms are assumed.
Here, the localization of molecules around a minority molecule leads to compartmentalization, whereas
infrequent replication of such a minority molecule induces the division of a compartment. Therefore,
the reproduction of a protocell, which is essentially a cluster of molecules, progresses in synchrony with the replication of
a minority molecule. We also demonstrate that this reproduction process tenaciously resists parasites.

We consider two molecular species that mutually catalyze the replication of each other. The reaction processes are expressed as follows:
\begin{equation}
X + Y \xrightarrow{p \gamma_X} 2X + Y, X+Y \xrightarrow{p \gamma_Y} , 2Y + X
\label{rep}
\end{equation}
\begin{equation}
X \xrightarrow{a_X} 0, Y \xrightarrow{a_Y} 0.
\label{decay}
\end{equation}
The first process represents mutual catalytic replication with rate $p$, where the synthesis fractions
of $X$ and $Y$ are $\gamma_X$ and $\gamma_Y$, respectively.
The second process represents the natural decomposition of $X$ and $Y$ with rates $a_X$ and $a_Y$, respectively.

Because it is necessary to ensure discreteness in a molecular number, we adopt stochastic simulations rather than
reaction-diffusion equations. In fact, the importance of ensuring discreteness in the molecular number has been discussed\cite{Togashi-KK, Solomon}.
We also take into account crowded molecules,
and we assume that both $X$ and $Y$ molecules have a spherical shape with the same diameter
($\sigma$). These molecules are confined to a box with dimensions $L_x \times L_y \times L_z$, with
elastic boundaries.
Further, these molecules have the same mass and diffusion constant ($D$).
Their motions obey the overdamped Langevin equations, which are numerically
integrated by the Ermak-McCammon algorithm\cite{ErmakMcCammon}, such that
\[ \Delta x = - \frac{D \Delta t}{k_B T} \nabla U(r) + S(t) \]
where $S(t)$ is a Gaussian white noise such that
\[ \langle S(t) \rangle = 0, \langle S(t)S(t') \rangle = 2D\Delta t \delta(t-t'). \]
 with temperature $k_B T = 1$, where $k_B$ denotes the Boltzmann constant.
In the simulation, we set $\Delta t = 0.00005$.
The interaction potential $U(r)$ between molecules is a Hertzian potential, $\phi(r) = E \left| \sigma - r \right|^{5/2} \Theta(\sigma - r)$,
where $E = 10000\epsilon/\sigma^{5/2}$ with an energy unit $\epsilon$, and the characteristic length scale $\sigma$ (i.e., the diameter) of molecules and $\Theta$ is a Heaviside step function\cite{note1}.
We start our simulation with the initial condition in which a single molecule $Y$ is surrounded by
$X$ molecules\cite{note2}.

At each time step, if the distance between every pair of $X$ and $Y$ is shorter than $\sigma$,
the replication reaction occurs with a probability $p' = p/\Delta t$.
In other words, $p$ denotes the rate of replicating molecules per unit time under a crowded condition,
i.e.,  when a pair of $X$ and $Y$ overlaps.
If the reaction occurs, a new molecule $X$ or $Y$ is added to the system,
with probability $\gamma_X$ or $\gamma_Y$, respectively.
Hence, the reaction rate for the replication of $X$ and $Y$ for a single reaction pair is given by $r_X \equiv p \gamma_X$ and $r_Y \equiv p \gamma_Y$, respectively.
The position of the new molecule is chosen as $(x, y, z) = (x_0 + f_x, y_0 + f_y, z_0 + f_z)$, 
where $(x_0, y_0, z_0)$ is the center between the reaction pair and
$(f_x, f_y, f_z)$, which is randomly chosen from $-\sigma < f_x, f_y, f_z < \sigma$, by also avoiding overlap with other molecules.
The decomposition process is carried out to remove
all $X$($Y$) molecules, with a probability $a'_X = a_X/\Delta t (a'_Y = a_Y/\Delta t)$ at each step.
Here, we fix the parameters as $\sigma = 1$, $D = 1$, and $p = 10^3$ and study the behavior of the system against the changes in
the independent parameters $\gamma_Y$, $a_X$, and $a_Y$, while $\gamma_X$ is set as $\gamma_X = 1 - \gamma_Y \geq \gamma_Y$ without loss of generality.

First, we examine the formation of a cluster by considering a single $Y$ molecule and suppressing its replication and decomposition
(i.e., $\gamma_Y = a_Y = 0$).
If $a_X$ is large, none of the $X$ molecules remain to exist, whereas
a stationary cluster of $X$ molecules is formed around the single $Y$ molecule
when $a_X$ is sufficiently small. 
For it, we fix $a_X = 4$ in the following.
Indeed, we have computed the density distribution function $N_X$ of $X$ as a function of
distance $r$ from the single $Y$.
The tail of the function for a large $r$ is fitted well by the isotropic steady-state solution of the diffusion equation $N_X \sim \exp(-r/r_0)/r$ with $r_0 = \sqrt{D/a_X}$.
For a small $r$, the value of the function is suppressed without $1/r$-divergence
owing to crowded molecules.
Under the crowded condition, $X$ and $Y$ continuously
overlap each other in the replication reaction.
Then, the replication rate of $X$ by a single $Y$ molecule, denoted as $R_X$,
is given by $R_X = r_X \overline{n}_X$, where $\overline{n}_X$ denotes the average number of $X$ molecules that overlap with the single $Y$ molecule.
In the present simulations, $\overline{n}_X \sim 2$\cite{note3}.
Under the present crowded condition,
the number of $X$ molecules, $N_{CX}$, in a steady cluster 
is estimated by the balance between
the replication and decomposition of $X$ molecules, as $R_X = a_X N_{CX}$.
By using $\overline{n}_X \sim 2$, we then estimate $N_{CX}$ as $\sim 500$, which is consistent with numerical results.

\begin{figure}
\includegraphics[height=5.5cm, clip]{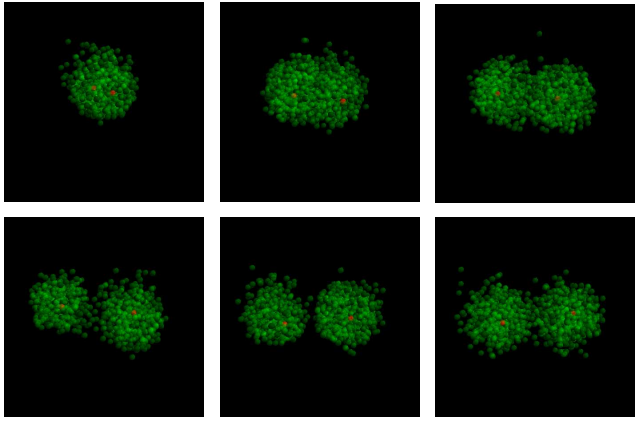}
\caption{Snapshots after the replication of $Y$. The green semitransparent particles represent the $X$ molecules. 
The red particles represent the $Y$ molecules, located deep within the clusters.
Time evolves from the left top to right bottom, as $t=0.1, 0.5, 1, 1.5, 2, 2.5$. The parameter values are
$r_Y = 0.01$, $a_X = 4$, and $a_Y = 0.002$. }
\label{fig:1}
\end{figure}

Now, we consider the replication of $Y$ molecules with a rate that is slower than that of $X$.
First, in Fig. \ref{fig:1}, we show an example of the growth-division process of a cluster and provide the condition for it later.
This process occurs as follows.
For most of the time, the cluster maintains itself by replicating the $X$ molecules, whereas the replication of the $Y$ molecule rarely occurs.
Once the $Y$ molecule replicates, the distance between the two $Y$ molecules is increased by diffusion, and these molecules contribute to
the replication of $X$ molecules.
As the two $Y$ molecules gradually move away from each other, the spatial distribution of $X$, synthesized by each $Y$, starts
to elongate in the direction of the two $Y$ molecules. Since the decomposition of the $X$ molecule is dominant because it is located
 away from a $Y$ molecule, the cluster of $X$ molecules forms a dumbbell shape, and it finally splits into two (see also supplementary movies in \cite{Movie}).

In fact, for this growth-division cycle to repeat, there is a restriction on the values of several parameters.
First, the replication of $Y$ should be sufficiently slow.
After the replication of a new $Y$ molecule, it takes some time for the two $Y$ molecules to be separated at a distance of the size of the cluster.
If another replication of $Y$ occurs during this time, the cluster of $X$ molecules cannot split into two because the replication of $X$
molecules by the third $Y$ molecule already begins.
Second, the decay rate of $Y$ should be sufficiently low. If this is not the case, the $Y$ molecules would completely disappear before the division of a cluster.
On the other hand, it should be noted that the cluster splits by the decomposition of $X$ so that its rate $a_X$ is higher than $a_Y$.

\begin{figure}
\rotatebox{270}{
\includegraphics[height=8cm, clip]{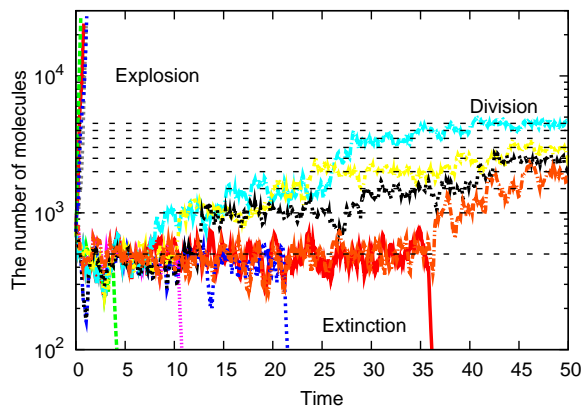}}
\caption{Time evolution of the total number of molecules. Data are obtained from the following three sets of parameters with several samples;
$r_Y = 0.001, a_Y =  0.2$ (leading to extinction), $r_Y = 0.01, a_Y = 0.002$ (division), and $r_Y = 1, a_Y = 0.02$ (extinction).
Here, we set $a_X = 4$. The dotted lines represent several multiples of the molecule number of a single cluster, $N_{CX}$. }
\label{fig:2}
\end{figure}

By taking several parameter values for $\gamma_Y$ and $a_Y$,
we have plotted the time evolution of the total number of molecules
by starting from a single-$Y$ initial condition in Fig. \ref{fig:2}.
We have found the following three typical behaviors: extinction, division, and explosion.
In the extinction phase, all the molecules decay, and finally no molecules remain.
In the parameters shown in this figure, extinction is initiated by the decomposition of a single $Y$ molecule,
followed by the extinction of $X$ molecules because there are no more replications.
Here, first, the number of molecules fluctuates at around a constant value, corresponding to a single cluster,
and then, it suddenly decreases with the disappearance of the $Y$ molecule.
(When $a_X$ is considerably larger than $a_Y$, the extinction of $X$ molecules precedes that of $Y$.)

\begin{figure}
\includegraphics[width=8cm, clip]{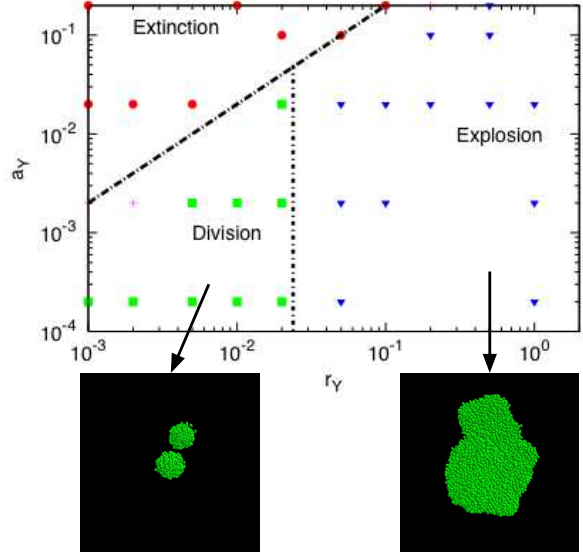}
\caption{Phase diagram of our model in $r_Y$-$a_Y$ space. The parameters are $p = 10^3$ and $a_X = 4$.
The plane is divided into the following three regions: extinction ($\circ$), division ($\square$), and explosion ($\triangledown$).
At the parameters with $+$ points, two behaviors of the adjacent regions coexist.
We also plot the theoretically estimated boundaries given by eqs. (\ref{b1}) and (\ref{b2}).}
\label{fig:3}
\end{figure}

In the division behavior, a stepwise increase in the total number of molecules is observed.
For reference, we also plot dotted lines in Fig.2 at the level of several multiples of the number of molecules in a single cluster.
The total number of molecules fluctuates around one of the dotted lines and gradually increases, as is
consistent with the division process of a certain size of a cluster.
In the explosion case, the number of molecules show an exponential increase without a step-like behavior.
The number of molecules increases by maintaining a single cluster.
Typical snapshots of the division and explosion cases are shown in Fig. \ref{fig:3}, where the green particles represent the $X$ molecules.

In Fig. \ref{fig:3}, a phase diagram of these three behaviors is plotted
on the $r_Y$-$a_Y$ plane by fixing the decay rate of $X$ at $a_X = 4$.
The three phases are separated in the parameter space.
The division phase lies at smaller $r_Y$ and $a_Y$.
An increase in $a_Y$ leads to extinction, whereas an increase in $r_Y$ leads to explosion.
Near the boundary between the division and the extinction phases, the corresponding two behaviors coexist,
depending on each sample.

The boundaries between the phases are estimated as follows.
  In the case of the boundary of the extinction phase, we note the balance between the replication and the decay of $Y$.
The replication rate of a $Y$ molecule in the cluster is estimated by $\sim R_Y = r_Y \overline{n}_X$,  and the decay rate is estimated by $\sim a_Y$.
In the case of the division phase, the replication rate of $Y$ should be greater than its decay rate 
so that the region satisfies
\begin{equation}
a_Y < R_Y.
\label{b1}
\end{equation}
This boundary is plotted in the $r_Y$-$a_Y$ plane, which is consistent with our simulation results.

The boundary between the division and the explosion phases is estimated by the balance between the replication rate of $Y$ and the size of the cluster.
The linear size of a single cluster, $L_C$, is estimated by $L_C \sim N_{CX}^{1/3}$.
The division process starts by the replication of $Y$ and is completed when the two $Y$ molecules diffuse approximately to the distance $L_C$.
We denote the typical timescale of dividing processes by $\tau_D$, and we obtain $L_C^2 \sim D_Y \tau_D$,
where $D_Y$ denotes the effective diffusion constant of $Y$.
Here, $D_Y$ is slightly increased from $D=1$ because of the repulsive interactions with $X$ molecules, and it is estimated as
$D_Y \sim 3$ from the simulations for Fig.3..

At the phase boundary, only a single replication of $Y$ should occur during this time $\tau_D$
to complete the division, so that $R_Y \tau_D = 1$.
Here, $\tau_D = L_C^2 / D_Y = (R_X / a_X )^{2/3}/D_Y$.
Thus,
the boundary is expressed as
\begin{equation}
R_Y = D_Y\left( \frac{a_X}{R_X}\right)^{2/3}.
\label{b2}
\end{equation}
The estimated boundary is plotted in Fig. 3, which is also consistent with our results.

In summary, we have studied a simple reaction system of two mutually catalytic molecules that diffuse under a crowded condition.
When the replication speed considerably differs between the two molecular species, the molecule with a slower replication rate becomes a minority molecule, 
and the minority molecule leads to the formation of a cluster of molecules and its division into two clusters as a result of replication of the molecule. This division is repeated to produce protocells of a certain size.
Here, we have presented the simulation results for a three-dimensional case; however, we have
confirmed the existence of
the present three phases, in particular, the division of a protocell
in a two-dimensional case.

To further examine whether the present protocell resists parasitic molecules,
we have also extended our model to include parasitic molecules $X'$ and $Y'$, which
are catalyzed by other molecular species but cannot catalyze the replication of other molecules.
If the variety and/or replication rates of the parasites are large, they would dominate.
Protocells dominated by parasites, however, stop growing and eventually disappear,
whereas cells that are not infected by parasites continue to grow and divide
(see Supplementary Material and Movie\cite{Movie}).
In particular, it is observed that when parasite $Y'$ appears, it is spontaneously emitted from the cluster, so that
$Y'$ parasites are eliminated. This is because $Y'$ cannot synthesize $X$; therefore, it cannot remain at
the center of the cluster, and it diffuses out of the cluster.

It should be noted that the reaction scheme given in eq. (\ref{rep}) is extremely simplified.
For such reactions to progress, 
some resources are necessary for the synthesis of $X$ and $Y$.
If such resources are sufficiently supplied, the original model (eq. (\ref{rep})) is derived. 
In reality, however, these resources have to be supplied from the outside\cite{Kamimura1}.
As the size of the cluster increases, their supply cannot be penetrated deeper into the cluster; therefore, replication is limited only at the periphery, whereas decomposition
occurs in a bulk.
Hence, the growth of a cluster without division stops when the cluster achieves a certain size.
In contrast, when the cluster divides into two before it reaches the size threshold,
the molecules continue to replicate.

In conclusion, the formation of a compartment and its growth-division cycle
are shown to be a natural outcome of mutually catalytic reactions with a minority molecule species.
In addition, this division of a protocell is synchronized with the replication of the minority molecule.
As already discussed, the minority molecule $Y$ can act as a carrier of heredity,
as it is preserved well, controls the characteristic of the protocells, and gives
evolvability\cite{KanekoYomo,Kaneko,Matsuura}.
Here, we have shown that the replication of this minority molecule is synchronized with the division of a protocell.
These characteristics of the minority molecule $Y$ agree well with those required by
genetic information, played by DNA in the present cell.
The question about the origin of genetic information from catalytic reactions, addressed by Dyson\cite{Dyson},
is thus answered.

This work is supported in part by the Japan Society for the Promotion of Science.

\newpage

\begin{center}
\section*{Supplementary Material for \\Reproduction of a Protocell by Replication of Minority Molecule \\in Catalytic Reaction Network}
\end{center}

In this Supplementary, we show that the protocell at the division phase in our model offers resistance to parasitic molecules,
which are synthesized with the help of other molecules, but cannot catalyze other molecules. Molecules $X$ and $Y$ are usually polymers, with a sequence of monomers.
Generally, there are errors in the replication of a (polymer) molecule; therefore, the sequence in $X$ or $Y$
can be modified. After the modification of the sequence, the molecules most likely lose
their catalytic activity, since such functional molecules are rather rare.
These molecules are often synthesized by other molecules, and thus, they are known as parasites.
Hence, the parasitic molecule $X'$ ($Y'$) replaces the original molecule
with a certain rate; this replacement process is regarded as the mutation of $X$ ($Y$).
In the replication processes of $X$ ($Y$), parasitic molecules such as $X'$ ($Y'$), and not $X$ ($Y$), are synthesized
with a probability $\mu$.
Because the sequence that has a catalytic activity for the synthesis of $Y$ ($X$)
 is rare, most mutants are parasites; therefore, backward mutations from $X'$ ($Y'$) to $X$ ($Y$) are neglected.

After the appearance of parasitic molecules, they replicate 
in the same way as the original $X$ and $Y$ molecules do.
\begin{equation}
X' + Y \xrightarrow{q} 2X' + Y (\gamma_X),
\label{p1}
\end{equation}
\begin{equation}
X' \xrightarrow{a_X} 0.
\end{equation}
\begin{equation}
Y'+ X \xrightarrow{q} 2Y' + X (\gamma_Y),
\label{p1}
\end{equation}
\begin{equation}
Y' \xrightarrow{a_Y} 0.
\end{equation}

In this supplementary, we assume that the replication and decay rates of parasitic molecules are
identical with the original molecules; however, it is observed that several simulations with different values give similar behaviors.
Time evolution is simulated from the initial condition, where a single $Y$ molecule is located at the center of
a few dozens of $X$ molecules.
When the mutation probability $\mu$ is considerably high, the parasitic
molecules dominate and they stop growing in some time.
When $\mu$ is not so high, the compartment splits into two before the parasites
accumulate completely, so that the compartment with a non-parasitic molecule pair continues to grow.

\begin{figure}
\begin{center}
\includegraphics[height=5cm, clip]{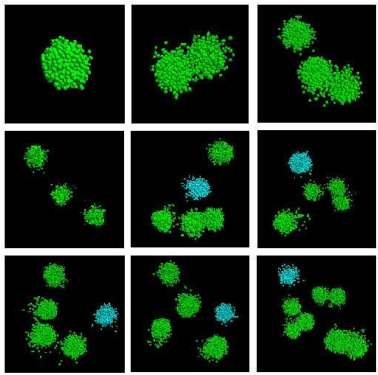}
\caption{Time evolution of the system with parasitic molecules. The green and light blue particles represent the $X$ and $X'$ molecules,
respectively. Snapshot patterns at $t = 5,10,15,20,30,35,40,45,50$ are plotted
from the left top to right, left middle to right middle, and left bottom to right bottom.
Compartments with $X$ molecules continue to grow whereas those covered by $X'$ do not grow.
Parameters are $p = q = 10^3, r_Y = 0.02, a_X = 4, a_Y = 2 \times 10^{-3}, \mu = 0.0001$.}
\label{sfig1}
\end{center}
\end{figure}

An example of temporal evolution is shown in Fig. \ref{sfig1}.
The number of molecules increase in time, and the cluster divides as in the case of the original model.
At some time (corresponding to the center middle box), parasitic molecules $X'$ appear by mutations and they replace the host molecule $X$.
Even if a compartment is dominated by parasites $X'$,
other compartments with non-parasitic pairs can continue to grow and divide.

\begin{figure}[htbp]
\begin{minipage}{0.45\hsize}
  \begin{center}
  \rotatebox{270}{
   \includegraphics[height=5cm, clip]{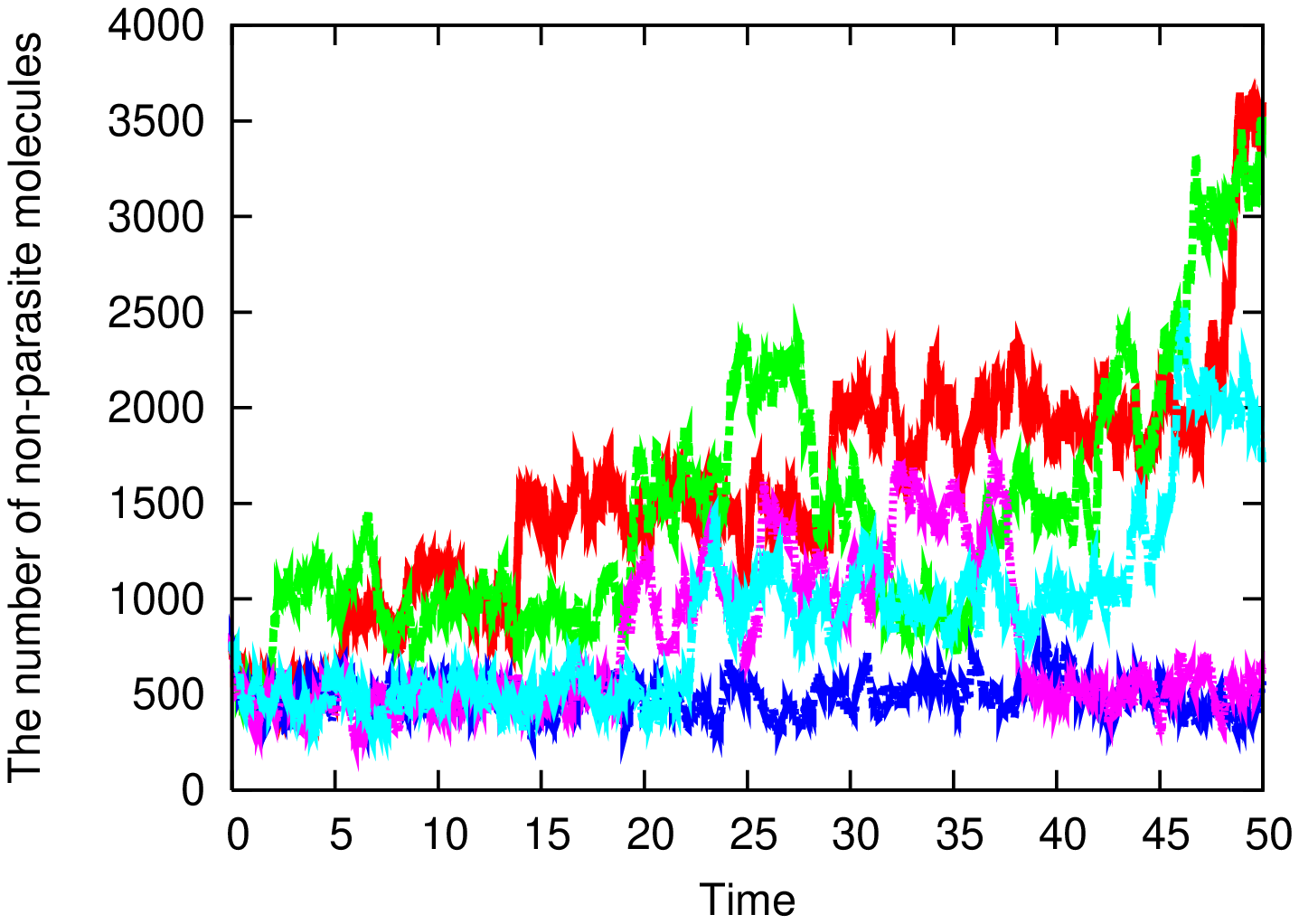}}
  \end{center}
  \caption{Time evolution of the number of non-parasitic molecules. Data are obtained for several samples, as plotted in different colors. The red line represents the sample shown in Fig. \ref{sfig1}.}
  \label{sfig2}
 \end{minipage}
 \begin{minipage}{0.45\hsize}
  \begin{center}
  \rotatebox{270}{
   \includegraphics[height=5cm, clip]{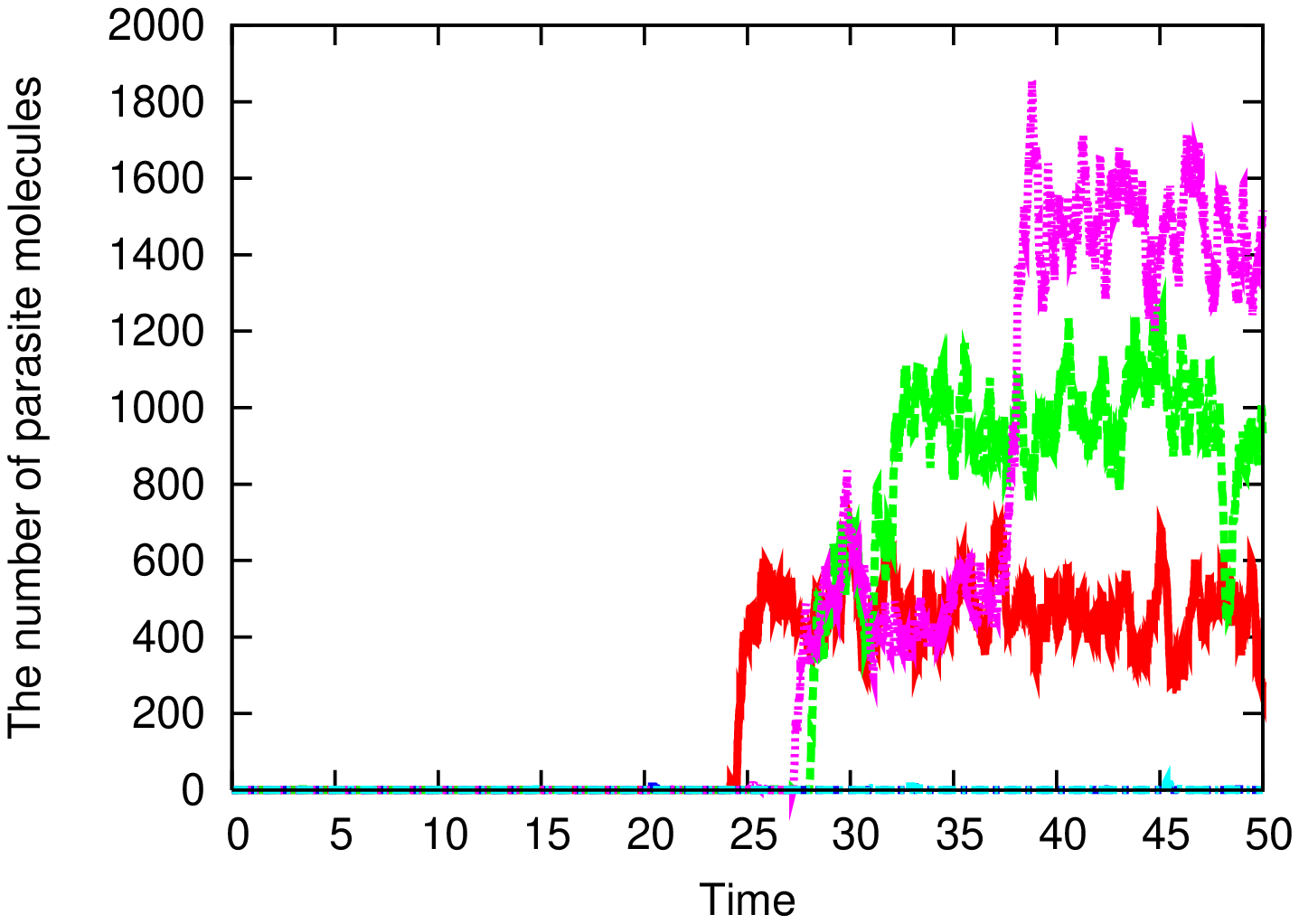}}
  \end{center}
  \caption{Time evolution of the number of parasitic molecules for the same set of
samples shown in Fig. \ref{sfig2}, plotted with corresponding colors.}
  \label{sfig3}
  \end{minipage}
\end{figure}

We have also plotted the time evolution of the number of non-parasitic and parasitic molecules for several samples in Figs. \ref{sfig2} and \ref{sfig3}, respectively.
In Fig. \ref{sfig2}, from the initial condition, the number of $X$ molecules fluctuates around the number of single clusters,
and some samples show a stepwise increase, as observed in our original model.
However, a sudden decrease in the number of $X$ molecules is also observed, and this decrease corresponds
to the appearance of parasitic molecules in Fig. \ref{sfig3}.
The number in Fig. \ref{sfig3} also shows stepwise increases corresponding to the number of a single cluster of $X'$.
These stepwise increases in $X'$ are caused by the transfer of non-parasitic pairs from some other cluster, 
and are not the autonomous growth by the replications of $Y$ within the cluster of the pair of $X'$ and $Y$.

\begin{figure}
\begin{center}
\includegraphics[height=5cm, clip]{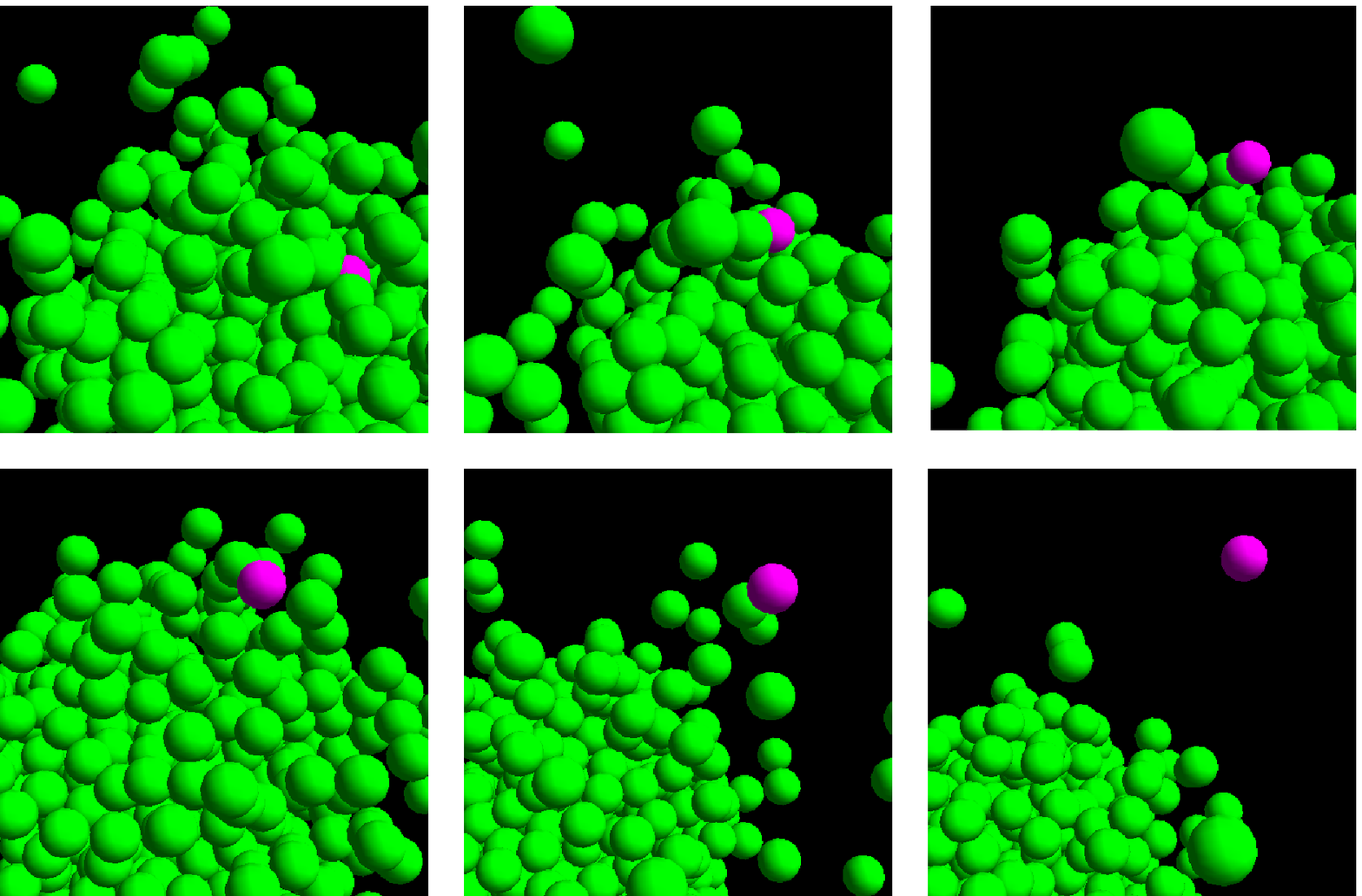}
\caption{Time evolution of the system with parasitic molecules $Y'$. The green and magenta particles represent the $X$ and $Y'$ molecules,
respectively. Snapshot patterns at $t = 1,1.5,2,2.5,3,3.5$ are plotted
from the left top to right and left bottom to right bottom.}
\label{sfig4}
\end{center}
\end{figure}

On the other hand, an example of resistance to the parasite $Y'$ is shown in Fig. \ref{sfig4}. A movie is also available in \cite{movie}. 
The $Y$ molecule is centered around a cluster of $X$ molecules since it
helps the synthesis of $X$ molecules. In contrast, parasitic $Y'$ molecules cannot aid the synthesis
of $X$; therefore, $X$ molecules do not gather in the vicinity of $Y'$ and $Y'$ freely diffuses until
it goes out of the cluster (see Fig. \ref{sfig4}).
Once the parasite $Y'$ diffuses out of the cluster,
it cannot replicate without the help of $X$, and therefore, it decomposes and eventually disappears.
Although we have assumed only mutual catalysis and diffusion of molecules, parasitic molecules
are selectively removed out of a cluster.

\begin{thebibliography}{999}
\bibitem{note1} Specific form is not important. Consideration of finite molecular size 
 is essential for the crowding effect.
\bibitem{note2} As long as there are $X$ and $Y$ molecules, the same behavior is exhibited after some time.
\bibitem{note3} This value depends on $D$ and $p$; $\overline{n}_X$ can be roughly written as $p\sigma^2/D$, whereas with a decrease in
$D$ it saturates at $\sim 12$, the value at the hexagonal close packing.
\end{thebibliography}
\end{document}